\documentclass[]{pasj02} 
\usepackage{url}
\usepackage{multirow}
\usepackage{ulem}
\usepackage{soul}

\jyear{2026}
\Received{}
\Accepted{}

\usepackage[usenames,dvipsnames]{xcolor}

\begin{document} 

\title{XRISM reveals sloshing-driven gas motions in the core of Abell 2029}

\author{
 Yuusuke \textsc{Uchida},\altaffilmark{1,2}\altemailmark\orcid{0000-0002-7962-4136}\email{uchida.yuusuke99@jaxa.jp} 
 Yuna \textsc{Saito},\altaffilmark{3}
 Naomi \textsc{Ota},\altaffilmark{4,5}\orcid{0000-0002-2784-3652}
 Erwin T. \textsc{Lau},\altaffilmark{4}\orcid{0000-0001-8914-8885}
 Ming \textsc{Sun},\altaffilmark{6}\orcid{0000-0001-5880-0703}
 Eric D. \textsc{Miller},\altaffilmark{7}\orcid{0000-0002-3031-2326}
 Tommaso \textsc{Bartalesi},\altaffilmark{8,9} \orcid{0009-0004-5838-2213}
 Stefano \textsc{Ettori},\altaffilmark{9,10}\orcid{0000-0003-4117-8617}
 Kotaro \textsc{Fukushima},\altaffilmark{11}\orcid{0000-0001-8055-7113}
 Caroline \textsc{Kilbourne},\altaffilmark{12}\orcid{0000-0001-9464-4103}
 Lorenzo \textsc{Lovisari},\altaffilmark{13,14}\orcid{0000-0002-3754-2415}
 Kyoko \textsc{Matsushita},\altaffilmark{11}\orcid{0000-0003-2907-0902}
 B.R. \textsc{McNamara},\altaffilmark{15}
 Arnab \textsc{Sarkar},\altaffilmark{16,7}\orcid{0000-0002-5222-1337}
 Kazunori \textsc{Suda},\altaffilmark{11}\orcid{0009-0003-3920-131X}
 and
 Irina \textsc{Zhuravleva}\altaffilmark{17}
}
\altaffiltext{1}{Institute of Space and Astronautical Science (ISAS), Japan Aerospace Exploration Agency (JAXA), Kanagawa 252-5210, Japan}
\altaffiltext{2}{Research Center for Space System Innovation, Research Institute for Science and Technology, Tokyo University of Science, 2641 Yamazaki, Noda, Chiba 278-8510, Japan}
\altaffiltext{3}{Faculty of Science and Technology, Tokyo University of Science, Chiba 278-8510, Japan}
\altaffiltext{4}{Department of Physics, Nara Women’s University, Kitauoyanishi-machi, Nara, Nara 630-8506, Japan}
\altaffiltext{5}{Argelander-Institut f\"{u}r Astronomie (AIfA), Universit\"{a}t Bonn, Auf dem H\"{u}gel 71, 53121 Bonn, Germany}
\altaffiltext{6}{Department of Physics and Astronomy, The University of Alabama in Huntsville, Huntsville, AL 35899, USA}
\altaffiltext{7}{Kavli Institute for Astrophysics and Space Research, Massachusetts Institute of Technology, Cambridge, MA 02139, USA}
\altaffiltext{8}{Dipartimento di Fisica e Astronomia “Augusto Righi” -- Alma Mater Studiorum -- Universit\`{a} di Bologna, I-40129 Bologna}
\altaffiltext{9}{INAF, Osservatorio di Astrofisica e Scienza dello Spazio, via Piero Gobetti 93/3, 40129 Bologna, Italy}
\altaffiltext{10}{INFN, Sezione di Bologna, viale Berti Pichat 6/2, 40127 Bologna, Italy}
\altaffiltext{11}{Department of Physics, Tokyo University of Science, 1-3 Kagurazaka, Shinjuku-ku, Tokyo 162-8601, Japan}
\altaffiltext{12}{NASA / Goddard Space Flight Center Greenbelt MD 20771, USA}
\altaffiltext{13}{INAF, Istituto di Astrofisica Spaziale e Fisica Cosmica di Milano, via A. Corti 12, 20133 Milano, Italy}
\altaffiltext{14}{Center for Astrophysics $|$ Harvard $\&$ Smithsonian, 60 Garden Street, Cambridge, MA 02138, USA}
\altaffiltext{15}{Department of Physics \& Astronomy, Waterloo Centre for Astrophysics, University of Waterloo, Ontario N2L 3G1, Canada}
\altaffiltext{16}{Department of Physics, University of Arkansas, 825 W Dickson st.,Fayetteville, AR 72701, USA}
\altaffiltext{17}{Department of Astronomy and Astrophysics, University of Chicago, 5640 S Ellis Ave, Chicago, IL 60637, USA}


\maketitle

\begin{abstract}
We investigate the velocity structure of the intracluster medium (ICM) in the core of the relaxed cool-core cluster Abell 2029 using XRISM Resolve spectroscopy. We analyze combined XRISM Resolve observations and divide the central region into several subregions. To account for photon mixing caused by the XRISM point spread function, we perform a spatial-spectral mixing analysis. We detect an ordered line-of-sight bulk-velocity gradient across the cluster core: the northern regions are blueshifted relative to the brightest cluster galaxy (BCG), while the southern regions are close to zero velocity or slightly redshifted. The maximum velocity difference is about $280~{\rm km\,s^{-1}}$. In contrast, the turbulent velocity dispersion is smaller, with measured values and upper limits of $\lesssim150~{\rm km\,s^{-1}}$, implying a non-thermal pressure fraction below $\sim2.5\%$. The velocity pattern is consistent with gas sloshing associated with the spiral structure seen in Chandra X-ray images. Averaged over all regions, the inferred turbulent heating rate is below the radiative cooling rate, indicating that turbulent dissipation alone is insufficient to offset cooling in the entire core. These results reveal that A2029 is not kinematically featureless: sloshing-induced bulk motions are present, while the observed line-of-sight velocity dispersion indicates only a limited contribution to pressure support and core heating.
\end{abstract}


\section{Introduction}

Galaxy clusters contain a hot X-ray-emitting intracluster medium (ICM) that traces the formation and evolution of the largest gravitationally bound structures in the universe. The dynamical state of the ICM provides important information about cluster assembly, energy transport, and the balance between heating and cooling in cluster cores. Gas motions in the ICM contribute to the non-thermal pressure support and transport kinetic energy, affecting both the thermodynamic structure of the gas and the accuracy of hydrostatic mass measurements (e.g., \cite{Lau09,Biffi16}).

Recent advances in high-resolution X-ray spectroscopy have enabled direct measurements of gas motions in the ICM \citep{Hitomi16,Hitomi_PSF18,Simionescu19,Sanders20}. Observations with the XRISM Resolve microcalorimeter \citep{Tashiro25,Ishisaki25,Kelley25} have shown that the velocity dispersion of the hot gas in several galaxy clusters is relatively small, indicating largely subsonic ICM motions. In particular, the cool-core cluster Abell 2029 \citep{Lewis02} exhibits a velocity dispersion of order $\lesssim150~{\rm km\,s^{-1}}$ and a small non-thermal pressure fraction in the central region and at intermediate radii \citep{XRISM25a,XRISM25b}. The line-of-sight bulk velocity of the hot gas is also small relative to the central galaxy. These results indicate that the ICM in A2029 is close to hydrostatic equilibrium on global scales, making it an excellent laboratory for studying subtle gas motions.

Despite this apparently quiescent state, deep Chandra observations of A2029 reveal a prominent spiral-like surface brightness structure extending to about 600~kpc \citep{Watson26,Paterno-Mahler13}. Such structures are widely interpreted as signatures of gas sloshing induced by minor mergers that perturb the cluster gravitational potential. Sloshing frequently produces spiral patterns and cold fronts in cool-core clusters \citep{Markevitch07,Ueda20} and is expected to drive large-scale bulk motions in the ICM \citep{ZuHone10,Roediger12}. The same observations further suggest that the spiral in A2029 is a relic of an off-axis minor merger several Gyr ago \citep{Watson26}. However, direct measurements of the associated velocity field have been difficult because of the limited spectral resolution of previous X-ray observatories.

Spatially resolved XRISM observations are now beginning to connect such X-ray structures with gas kinematics. In the Centaurus cluster, XRISM detected bulk flows of $130$ to $310~{\rm km\,s^{-1}}$ consistent with sloshing, while the velocity dispersion remains relatively small even near the central active galactic nucleus \citep{XRISM25d}. In the Perseus cluster, XRISM revealed multiple kinematic drivers on different spatial scales, with active galactic nucleus feedback dominating the inner core and merger-driven motions likely associated with sloshing dominating the outer region \citep{XRISM26a,Zhang26}. These results highlight the importance of spatially resolved velocity measurements for identifying the drivers of gas motions in cluster cores.

A2029 provides a useful opportunity to investigate the coexistence of a globally quiescent ICM and a large-scale sloshing spiral. If the spiral is produced by sloshing, coherent line-of-sight velocity gradients should be present even when the non-thermal pressure support remains small. Spatially resolved velocity measurements therefore provide a direct kinematic test of the sloshing interpretation inferred from Chandra imaging.

In this paper we carry out such a test using spatially resolved spectroscopy with XRISM Resolve. We subdivide the A2029 core to map the spatial distributions of bulk velocity and velocity dispersion, and we further use the measured velocity dispersions to estimate the turbulent heating rate and compare it with the radiative cooling rate in the cool core.

The cosmological parameters are $\Omega_{m0}=0.3$, $\Omega_{\Lambda}=0.7$, and $h=0.7$ throughout this paper. Accordingly, $1\arcmin$ corresponds to 89~kpc at the redshift of A2029 ($z=0.0787$). We report redshifts and velocities corrected to the Solar System barycenter. We use the proto-solar abundance table from \citet{Lodders09}. The quoted errors represent the $1\sigma$ statistical uncertainties unless stated otherwise.

This paper is organized as follows. Section~\ref{sec:observation} describes the XRISM observations and data reduction. Section~\ref{sec:analysis} presents the spectral analysis and measurements of the gas velocity structure. Section~\ref{sec:discussion} discusses the implications for sloshing, non-thermal pressure support, and the heating and cooling balance. Section 5 summarizes the main results.

\section{Observations and data reduction}\label{sec:observation}

\subsection{XRISM observations}

We analyzed XRISM Resolve observations of the galaxy cluster Abell 2029 obtained during the performance verification phase and the subsequent GO program. The performance verification data include two observations of the central region on 2024 January 10 and 13 (OBSID 000149000 and 000151000) and one northern offset pointing, N1, observed on 2024 January 10 (OBSID 000150000). The GO data include an additional central observation obtained on 2025 July 17 (OBSID 201042010). The cleaned exposure times are listed in Table~\ref{tab:obslog}. The spatially resolved subarray analysis presented below uses the combined data set, including the GO center observation to improve the photon statistics of the core regions.

Barycentric velocity corrections were applied to all Resolve observations. The corrections are approximately $+26~{\rm km\,s^{-1}}$ for the January 2024 observations and $-27~{\rm km\,s^{-1}}$ for the July 2025 observation.

\begin{table*}[htb]
  \centering
  \tbl{Observation information for each pointing}{
  \begin{tabular}{llllll}
    \hline\hline
    Object & OBSID & Observation start & Exposure time [ks] & RA\_NOM [deg] & DEC\_NOM [deg]\\ 
    \hline
    Abell2029\_Center & 000149000 & 2024-01-10T04:03:46 & 15.2 & 227.7341 & 5.7452\\
    Abell2029\_N1 & 000150000 & 2024-01-10T16:51:55 & 90.6 & 227.7635 & 5.7850\\
    Abell2029\_Center & 000151000 & 2024-01-13T01:50:41 & 26.9 & 227.7332 & 5.7451\\
    Abell2029\_Center & 201042010 & 2025-07-17T12:51:00 & 118.3 & 227.7330 & 5.7445 \\ \hline
  \end{tabular}}\label{tab:obslog}
\end{table*}

\subsection{Data reduction}
\label{sec:data_reduction}

The analysis started from the cleaned event files processed with pipeline version 03.00.013.009. In addition to the standard screening criteria, we applied the recommended screening in the XRISM ABC Guide. We extracted spectra using only high primary grade events (ITYPE == 0), which provide the highest spectral resolution for Resolve. The calibration pixel (pixel 12) and pixel 27 were excluded because pixel 27 has exhibited gain variations distinct from other pixels.

\begin{figure*}[tbh]
\begin{center}
\includegraphics[width=0.45\linewidth]{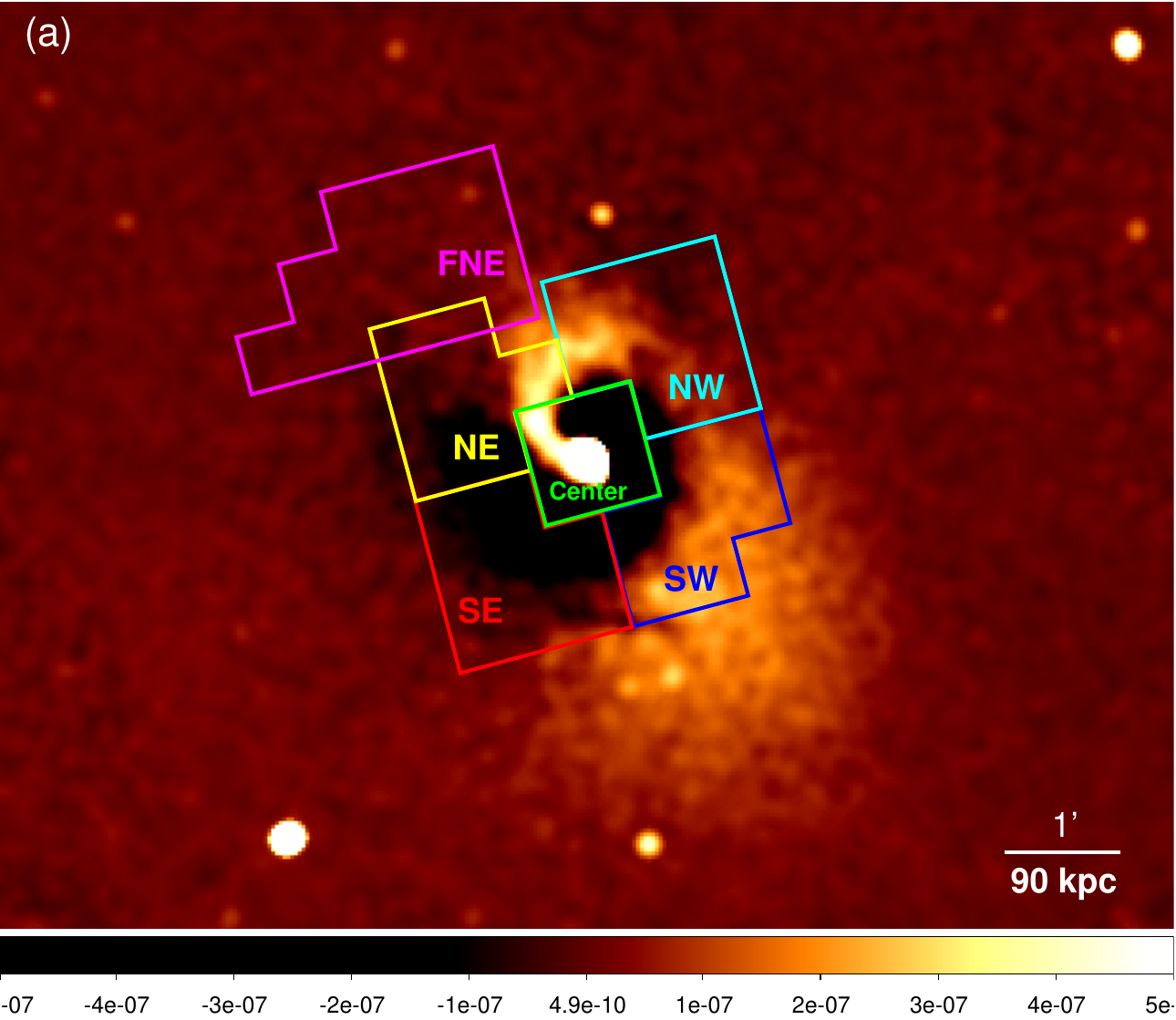}
\includegraphics[width=0.45\linewidth]{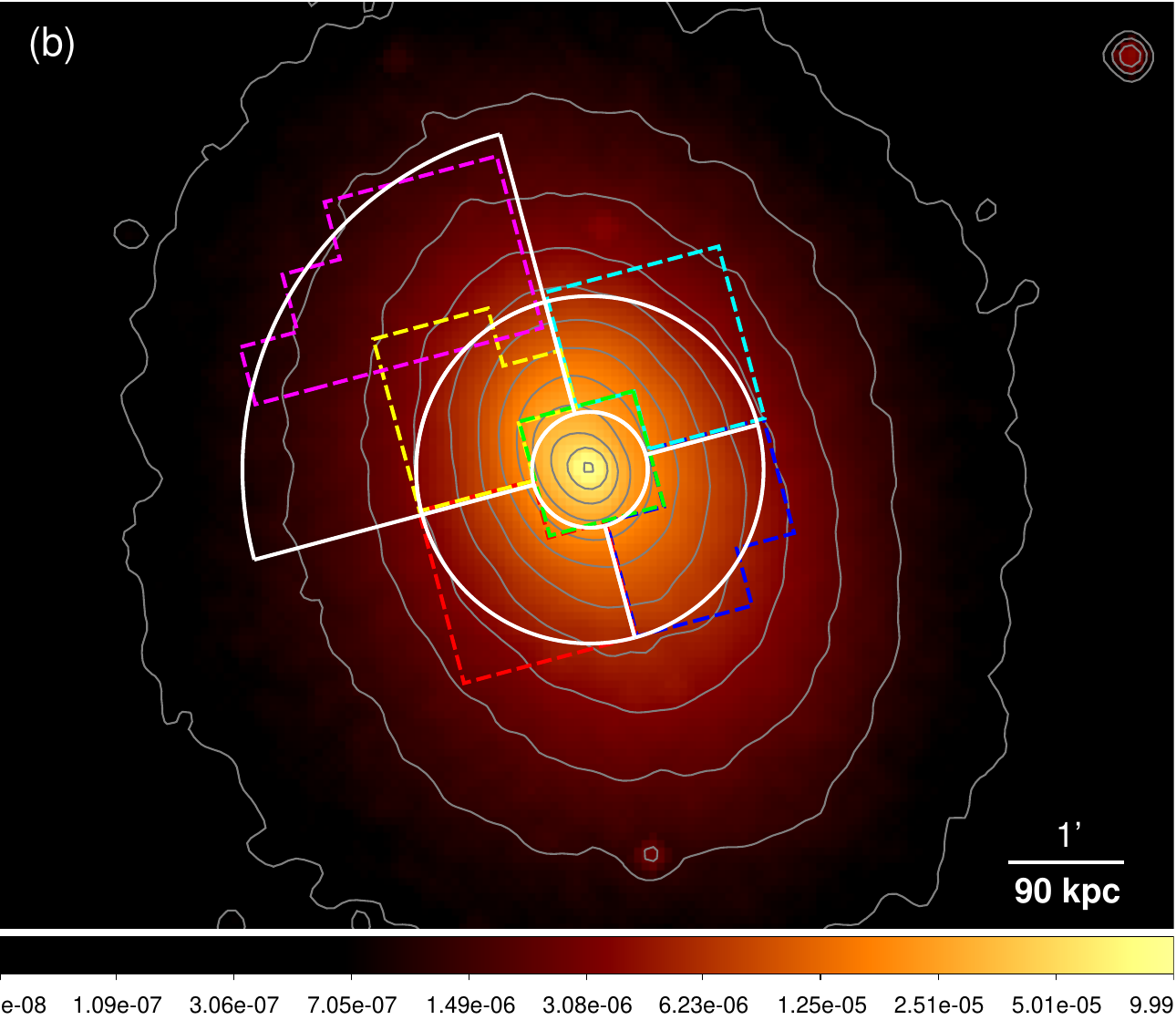}
\end{center}
\caption{(a) Chandra residual X-ray image of the A2029 core shown in linear color scale. The image is smoothed with a Gaussian kernel of $\sigma = 4$~arcsec. The spectral extraction regions used in this work are indicated by dashed outlines and labeled as center, NE, NW, SE, SW, and FNE. (b) Region definition used for ARF generation. The Chandra image is divided into a central circular region with a radius of 0.5 arcmin and two concentric annuli covering 0.5-1.5 arcmin and 1.5-2.5 arcmin. The annuli are further subdivided into four azimuthal sectors, resulting in six source regions. Gray contours show the X-ray surface brightness in logarithmic spacing with levels starting from $1\times10^{-7}$ and increasing to $1\times10^{-4}$ in units of ${\rm counts\,s^{-1}cm^{-2}pixel^{-1}}$. Dashed lines indicate the spectral extraction regions shown in panel (a).
{Alt text: Two panel X-ray images of the A2029 core.
}}\label{fig:region}
\end{figure*}

For spatially resolved spectroscopy, we defined six subregions as shown in Figure~\ref{fig:region}. The central pointing was divided into five regions: center, north east (NE), north west (NW), south east (SE), and south west (SW). We also defined a far north east (FNE) region using the N1 pointing to represent emission from the outer part of the cluster and evaluate contamination from outside the central field of view.

The redistribution matrix files (RMFs) were generated for each region using rslmkrmf in HEASoft version 6.34 with CalDB v20250315. We adopted the large-size RMF generated with the option whichrmf = L.

The ancillary response files (ARFs) were produced using \texttt{xaarfgen} in HEASoft, with a Chandra X-ray image as the input surface brightness distribution. This accounts for the spatial distribution of cluster emission within the Resolve field of view. The treatment of spatial-spectral mixing caused by the XRISM point spread function is described in Section~\ref{sec:analysis}.

The spectral analysis focuses on the energy band above 2~keV, where calibration uncertainties associated with the gate valve transmission are minimized. The systematic uncertainty in the Resolve energy scale at 6 keV is approximately 0.3 eV \citep{Eckart25}, corresponding to about $15~{\rm km\,s^{-1}}$, and the instrumental line-width uncertainty corresponds to approximately $3~{\rm km\,s^{-1}}$. These calibration uncertainties are included in the quoted velocity uncertainties.

\section{Analysis and results}\label{sec:analysis}

Our main goal is to measure the gas kinematics in the central region of Abell 2029 using the combined performance-verification (PV) and General Observer program (GO) data set. Previous studies by \citet{XRISM25a} and \citet{XRISM25b} reported field-averaged thermodynamic properties using the full Resolve array, while here we focus on the spatially resolved velocity structure in the cluster core. Because the point spread function of XRISM is comparable to the Resolve field of view, spatially resolved spectroscopy requires careful treatment of photon mixing between regions. We therefore perform the spectral analysis using the spatial-spectral mixing method. This section describes the spectral modeling, implementation of the spatial-spectral mixing analysis, and resulting velocity measurements.

\subsection{Spectral modeling}

We performed spectral modeling and fitting for the Resolve data with XSPEC version 12.14.1 \citep{Arnaud96}. The ICM emission was modeled using the thermally broadened collisional ionization equilibrium model with velocity broadening, bapec \citep{Smith01}, based on AtomDB version 3.0.9 \citep{Foster12}. Galactic absorption was represented by TBabs \citep{Wilms00}, with a column density fixed at $N_{\rm H}=3\times10^{20}$ cm$^{-2}$ \citep{HI4PI16}. In the central region analyzed here, the cosmic X-ray background is negligible compared with the cluster emission and is therefore not included in the spectral modeling \citep{XRISM25b,Sarkar25}. The spectra were grouped using \texttt{ftgrouppha} to have at least one count per bin, and the fitting was performed by reducing the C-statistic \citep{Cash79}. 

\begin{figure}[htb]
\begin{center}
\includegraphics[width=0.45\textwidth,clip]{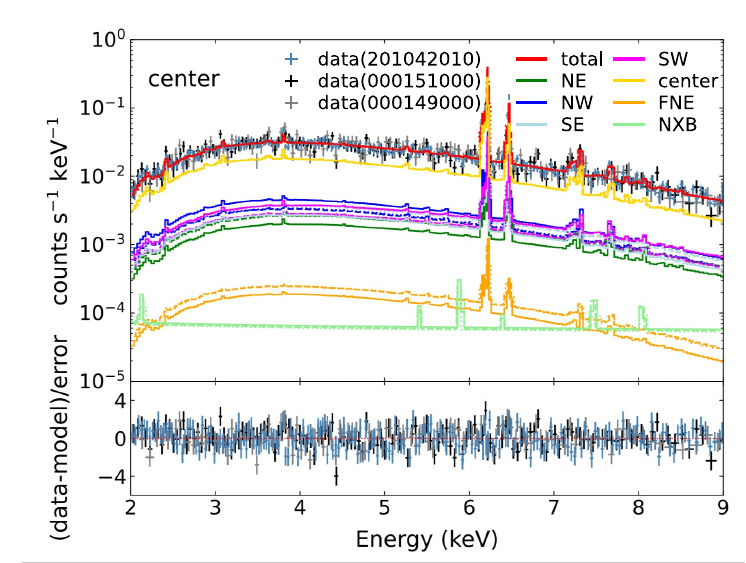}
\end{center}
\caption{Resolve spectrum of the center region and the best-fit spectral model. Blue, gray, and black crosses represent the spectra from observation IDs (OBSIDs) 201042010, 000151000, and 000149000, respectively. The red curve shows the total best-fit model. Individual model components are also shown, including contributions from the center (yellow), NE (green), NW (blue), SE (cyan), SW (magenta), and FNE (orange) regions due to PSF mixing, together with the non-X-ray background component (light green). The solid line represents the best-fit model for GO-observation, while the dahsed and the dotted lines represent the best-fit model for PV-observations.
The results for the other regions are shown in Figure~\ref{fig:app_spectra_all}. 
{Alt text: X-ray spectrum of the A2029 center region. The x-axis shows the energy from 2 to 9 keV. The y-axis shows the counts per second per keV.
}}
\label{fig:spec_center}
\end{figure}

The non-X-ray background (NXB) spectra were generated for each region using \texttt{rslnxbgen} with the archival Resolve night Earth database version 2\footnote{The NXB generation procedure is documented at \url{https://heasarc.gsfc.nasa.gov/docs/xrism/analysis/nxb/resolve_nxb_db.html}.}. 
For each region, we generated the background spectrum using \texttt{rslnxbgen} and reproduced it with a model consisting of a power-law continuum and Gaussian components representing instrumental fluorescent lines. The normalization of the background model was determined by fitting the reproduced background spectrum and was then fixed in the subsequent ICM spectral analysis. An example of the Resolve spectrum and best-fit model for the center region is shown in Figure~\ref{fig:spec_center}.

\subsection{Spatial spectral mixing analysis}

To estimate photon contamination from neighboring regions, we performed a spatial-spectral mixing (SSM) analysis. This contamination was primarily due to the XRISM Resolve XMA point spread function (PSF) with a half-power diameter of 1.3 arcmin. To account for this effect, we followed the SSM procedure adopted in recent XRISM studies of galaxy clusters \citep{XRISM25b,XRISM25d}.

We assumed that the incoming photons originated from six sky regions, corresponding to the six Resolve regions. We defined the sky regions as a central circular region with a radius of 0.5 arcmin and two concentric annuli covering radii of 0.5--1.5 arcmin and 1.5--2.5 arcmin. The inner annulus was divided into four azimuthal sectors. For the outer annulus, only the northeastern sector was included. This configuration resulted in a total of six source regions (Figure~\ref{fig:region}(b)). For each pair of source and extraction regions, we generated an ARF describing the fraction of photons from source regions detected in the corresponding extraction region. We prepared 88 ARFs\footnote{The central pointing observation consisted of three datasets. For Center, NE, NW SE and SW, the total number of source--extraction region ARF was $3\times5\times5=75$. Since spillover from Center, NE and NW regions to FNE region was significant, the related ARFs were $3\times3=9$. We also generated ($3 \times 1 = 3$) ARFs to account for spillover from the FNE region into the Center, NE, and NW regions, because the FNE region had only one dataset. Including the self-ARF for the FNE region, the total number of ARFs was 88.} for all combinations of the six source regions except for the combinations of FNE$\leftrightarrow$SW and FNE$\leftrightarrow$SE because the PSF spillover fraction between those two pairs is at most about one percent.

The spectra from all regions were fitted simultaneously using the multiple-response functionality of XSPEC, in which each observed spectrum is modeled as a linear combination of emission from all source regions weighted by the corresponding ARFs. The best-fit spectral parameters obtained from the BAPEC model are summarized in Table~\ref{tab:spectral_results}.

\begin{table*}[htb]
\tbl{Spectral fitting results of the BAPEC model for the six regions and derived non-thermal pressure fractions.}{
\begin{tabular}{lllllllll}
\hline\hline
Region & $kT$ & $Z$ & $z$$^a$ & $v_{\rm bulk}$ & $\sigma_v$ & norm$^b$  & $\alpha$$^c$ & $\alpha_{\rm turb}$$^c$\\
 & (keV) & (solar) &  & (km s$^{-1}$) & (km s$^{-1}$) & (10$^{-4}$ cm$^{-5}$) & &  \\
\hline
FNE    & $6.52_{-0.21}^{+0.23}$ & $0.30_{-0.03}^{+0.03}$ & $0.077214_{-0.000171}^{+0.000053}$ & $-191_{-50}^{+21}$ & $90_{-59}^{+35}$ & $0.806\pm0.020$ & $1.94_{-1.16}^{+0.65}$ & $0.79_{-1.02}^{+0.61}$ \\
NE     & $8.74_{-0.36}^{+0.57}$ & $0.78_{-0.07}^{+0.11}$ & $0.077636_{-0.000068}^{+0.000081}$ & $-73_{-24}^{+27}$  & $0_{-0}^{+32}$ & $0.495_{-0.021}^{+0.018}$ & $0.13_{-0.08}^{+0.09}$ & $<0.07$ \\
NW     & $6.55_{-0.15}^{+0.16}$ & $0.54_{-0.03}^{+0.03}$ & $0.077244_{-0.000059}^{+0.000049}$ & $-183_{-22}^{+20}$ & $123_{-18}^{+17}$ & $0.989_{-0.017}^{+0.018}$ & $2.48_{-0.47}^{+0.44}$ & $1.45_{-0.41}^{+0.39}$ \\
Center & $5.27_{-0.23}^{+0.15}$ & $0.74_{-0.05}^{+0.03}$ & $0.078050_{-0.000052}^{+0.000036}$ & $42_{-21}^{+18}$   & $131_{-21}^{+16}$ & $1.369_{-0.035}^{+0.047}$   & $2.11_{-0.64}^{+0.48}$ & $2.04_{-0.64}^{+0.48}$ \\ 
SE     & $7.81_{-0.23}^{+0.31}$ & $0.43_{-0.04}^{+0.03}$ & $0.078225_{-0.000048}^{+0.000093}$ & $90_{-20}^{+30}$   & $74_{-74}^{+40}$ & $0.752_{-0.016}^{+0.015}$ & $0.66_{-0.88}^{+0.50}$ & $0.44_{-0.88}^{+0.48}$ \\
SW     & $8.28\pm0.25$ & $0.60\pm0.04$ & $0.077940_{-0.000075}^{+0.000039}$ & $11_{-26}^{+18}$   & $80_{-36}^{+26}$ & $0.809\pm0.015$ & $0.49_{-0.44}^{+0.31}$ & $0.49_{-0.44}^{+0.31}$ \\
\hline
\end{tabular}}
\label{tab:spectral_results}
\begin{tabnote}
The best-fit statistic is C-stat/d.o.f. = 20241.01/22134.
$^a$ The redshift values are corrected for the barycentric motion of the Earth. 
$^b$ Normalization of the BAPEC model defined as $(10^{-14}/4\pi [D_A(1+z)]^2)\int n_e n_H dV$. 
$^c$ $\alpha$ and $\alpha_{\rm turb}$ denote the non-thermal pressure fractions (in \%) estimated from both bulk and turbulent motions, and from turbulent motions only, respectively. 
\end{tabnote}
\end{table*}

\subsection{Velocity measurements}\label{subsec:analysis_velocity}

Using the spectral modeling and spatial-spectral mixing analysis described above, we derived the temperature, metal abundance, bulk velocity, and turbulent velocity in each region. From the redshift $z$ obtained in each region and the BCG redshift $z_{\rm BCG}=0.0779$ \citep{XRISM25a}, the line-of-sight bulk velocity relative to the BCG is calculated as

\begin{equation}
v_{\rm bulk}=\frac{c(z-z_{\rm BCG})}{1+z_{\rm BCG}} .
\end{equation}

The sound speed in each region is derived from the fitted temperature as $c_{\rm s}=\sqrt{\gamma k_{\rm B}T/\mu m_{\rm p}}$, where $\gamma=5/3$ is the adiabatic index, $k_{\rm B}$ is the Boltzmann constant, $\mu=0.61$ is the mean molecular weight, and $m_{\rm p}$ is the proton mass. Assuming isotropic turbulence, the turbulent Mach number is
\begin{equation}
\mathcal{M}_{\rm 3D}=\frac{\sqrt{3}\sigma_v}{c_{\rm s}},
\label{eq:Mturb}
\end{equation}
where $\sigma_v$ is the one-dimensional turbulent velocity dispersion. The corresponding non-thermal pressure fraction from turbulent motions alone is
\begin{equation}
\alpha_{\rm turb}=\frac{\mathcal{M}_{\rm 3D}^2}{\mathcal{M}_{\rm 3D}^2+3/\gamma}.
\label{eq:fnt_turb}
\end{equation}
Following \citet{XRISM25b}, we also define the effective three-dimensional Mach number including both turbulent and bulk motions as
\begin{equation}
\mathcal{M}_{\rm 3D,eff}=\frac{\sqrt{3\sigma_v^2+v_{\rm bulk}^2}}{c_{\rm s}},
\label{eq:Meff}
\end{equation}
and the corresponding total non-thermal pressure fraction as
\begin{equation}
\alpha=\frac{P_{\rm NT}}{P_{\rm tot}}=\frac{\mathcal{M}_{\rm 3D,eff}^2}{\mathcal{M}_{\rm 3D,eff}^2+3/\gamma}.
\label{eq:fnt}
\end{equation}

The SSM analysis of the combined data set is adopted as the baseline result below. Comparisons between analyses with and without SSM, with the temperature fixed to Chandra values, and between data subsets are used to evaluate systematic uncertainties and are described in Appendix~\ref{app:systematics}. We also examine the possible effect of resonant scattering by repeating the fits after excluding the Fe~\textsc{xxv} resonance line. These checks show that the bulk velocity pattern is robust, while the turbulent velocity is more sensitive to PSF mixing and to the treatment of the strongest Fe line.

Compared to the field-averaged measurements reported by \citet{XRISM25a}, the temperature shows clear spatial variations among the six regions. The center and FNE regions are cooler than $6.83_{-0.13}^{+0.14}$ keV, the NE, SE, and SW regions are hotter, and the NW region is consistent with this value within uncertainties. The metal abundance also varies spatially, with relatively high values in the center and NE regions and lower values in the SE and FNE regions, rather than being uniform around $0.60_{-0.02}^{+0.02}$ solar.

\begin{figure*}[htb]
\begin{center}
\includegraphics[width=0.96\textwidth,clip]{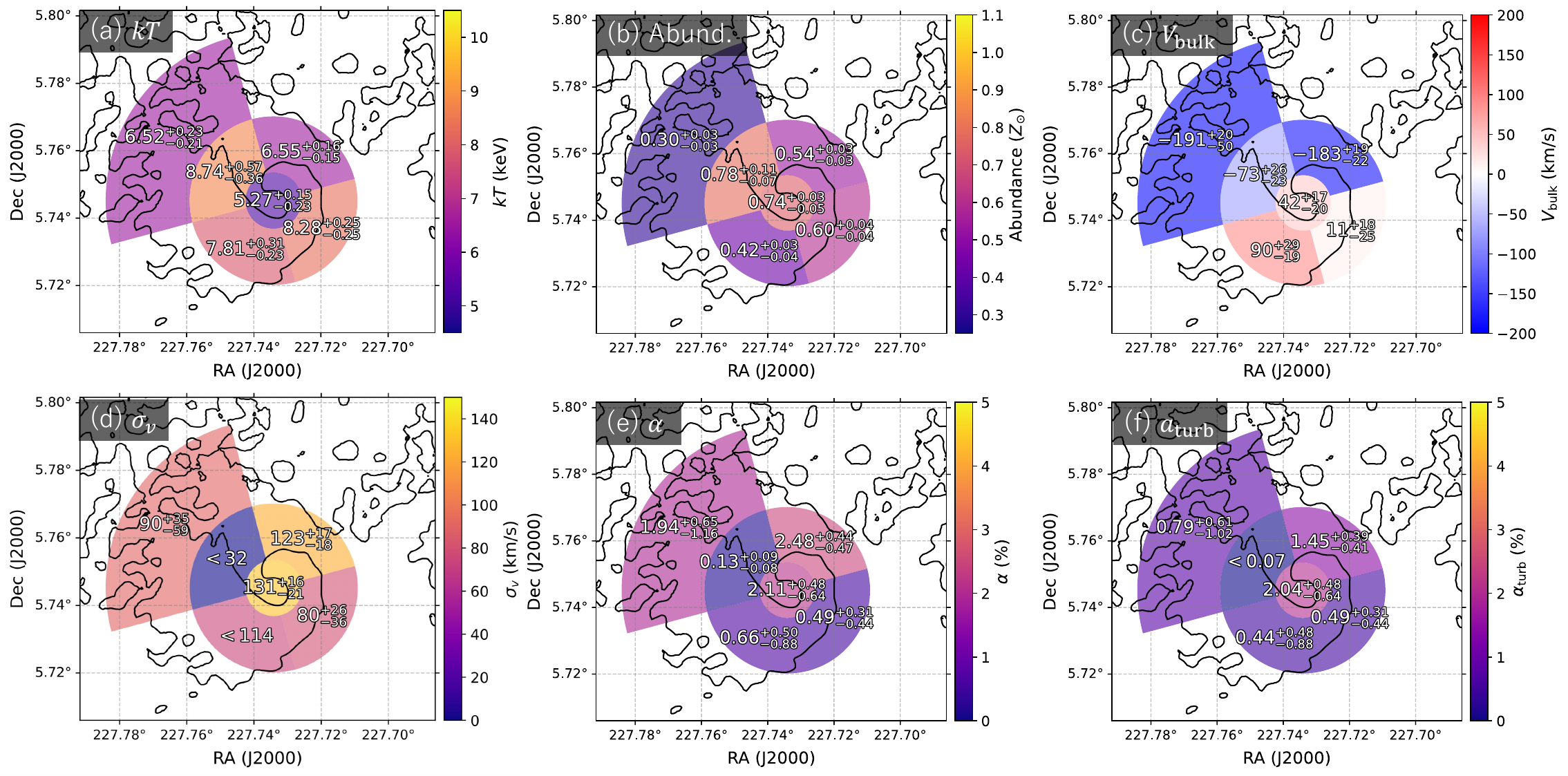}
\end{center}
\caption{Spatial distribution of the best-fit parameters in the cluster core derived from the spatial spectral mixing analysis. Panels show (a) temperature, (b) metal abundance, (c) bulk velocity relative to the BCG, (d) turbulent velocity dispersion, (e) non-thermal pressure fraction, and (f) non-thermal pressure fraction from turbulent motions alone. The color scale in each panel represents the best-fit value of the corresponding parameter. The numbers shown in each sector indicate the best-fit values with their $1\sigma$ statistical uncertainties. Black contours show the 0 level of the Chandra residual X-ray image shown in Figure~\ref{fig:region}.
{Alt text: Six parameter maps of the A2029 core using the same sector regions.
}}
\label{fig:par_map}
\end{figure*}

Negative bulk velocities are observed in the NE, NW, and FNE regions, indicating blueshifted motions relative to the BCG. The FNE region shows the largest blueshift of approximately $-190~{\rm km\,s^{-1}}$. In contrast, the SE and SW regions have velocities close to zero or slightly positive, reaching about $+90~{\rm km\,s^{-1}}$. The maximum velocity difference, between the FNE and SE regions, is therefore approximately $280~{\rm km\,s^{-1}}$. To quantify the north-south velocity gradient, we computed the inverse-variance weighted mean velocities of the northern regions (NE, NW, and FNE) and the southern regions (SE and SW). We obtain $\langle v_{\rm bulk}\rangle_{\rm N}=-151.0\pm11.3~{\rm km\,s^{-1}}$ and $\langle v_{\rm bulk}\rangle_{\rm S}=42.4\pm12.3~{\rm km\,s^{-1}}$, corresponding to $\Delta v=193.3\pm16.7~{\rm km\,s^{-1}}$ ($11.5\sigma$). Even after adding the current Resolve energy-scale systematic uncertainty of $15~{\rm km\,s^{-1}}$ in quadrature to the velocity uncertainty of each region, the significance remains $8.8\sigma$. Thus, the north-south velocity gradient is robust against the known calibration uncertainty. The non-thermal pressure fraction is slightly higher in the central and northern regions than in the southern regions but remains below 2.5\% in all regions. The spatial distributions of the best-fit parameters are summarized in Figure~\ref{fig:par_map}.

\section{Discussion}
\label{sec:discussion}
We discuss the physical implications of the velocity structure measured in the core of A2029. We first interpret the observed line-of-sight velocity gradient as sloshing-induced gas motion and compare it with numerical simulations. We then examine the small non-thermal pressure support inferred from the velocity dispersion and discuss the possible role of gas motions in the heating and cooling balance of the cool core.

\subsection{Sloshing-induced gas motions in the core of A2029}
\label{sec:discussion_sloshing}

A key result of this work is the detection of a coherent north-south line-of-sight velocity gradient across the core of A2029. As reported in Section~\ref{subsec:analysis_velocity}, Table~\ref{tab:spectral_results}, and Figure~\ref{fig:par_map}, the northern regions are blueshifted relative to the BCG, whereas the southern regions are consistent with zero velocity or slightly redshifted, giving a maximum contrast of $\Delta v_{\rm bulk} \sim 280~{\rm km\,s^{-1}}$ across the central few hundred kiloparsecs. This systematic pattern indicates that the hot gas is not completely at rest in the cluster potential, even though A2029 has often been regarded as a dynamically relaxed cool-core cluster.

Deep Chandra observations have revealed a large-scale spiral in the ICM of A2029, extending several hundred kiloparsecs from the center and interpreted as the imprint of a past off-axis minor merger \citep{Watson26}. The velocity gradient measured with XRISM is naturally explained in this context: the opposite signs of $v_{\rm bulk}$ between the northern and southern regions indicate a coherent line-of-sight component of the same gas motion. This sign change is difficult to ascribe to a uniform calibration offset, and the same north--south pattern is recovered in the independent without-SSM fits (Appendix~\ref{app:systematics}, Figure~\ref{fig:app_model_comparison}), confirming that it is not an artifact of the mixing correction.

The velocity contrast should not be interpreted as the velocity of a single gas element, but rather as the difference in emission-weighted projected velocities between different parts of the core. The observed line-of-sight bulk velocities are well below the adiabatic sound speed of $c_{\rm s} \sim 1300$--$1500~{\rm km\,s^{-1}}$ for gas temperatures of $kT \sim 7$--$8~{\rm keV}$, corresponding to bulk Mach numbers of only $\mathcal{M}_{\rm bulk} \sim 0.1$--$0.15$. Even using the full velocity difference as a characteristic amplitude, the motion remains clearly subsonic, indicating gentle bulk motion rather than a violent merger or shock-dominated flow.

The observed velocity gradient supports the interpretation that the spiral structure in A2029 is accompanied by coherent gas motions, although a one-to-one spatial correspondence between the Chandra surface-brightness residuals and the XRISM velocity extrema is not expected. This picture is also consistent with the thermodynamic and chemical structure reported in previous studies. The Chandra data show that the inner spiral excess is associated with relatively cool gas and enhanced metallicity, as expected if low-entropy, metal-rich core gas has been displaced from the center \citep{Watson26}. In addition, the XRISM multi-temperature analysis found cooler gas components in the central and inner northern regions, possibly tracing the same displaced cool gas \citep{Sarkar25}. We compare this interpretation with numerical simulations in the next section.

\subsection{Comparison with numerical simulations of sloshing}
\label{sec:discussion_simulation}

Idealized simulations of off-axis minor mergers show that a perturbing subcluster can displace the cool core from the bottom of the gravitational potential and excite sloshing motions, producing spiral cold fronts and organized gas flows \citep{ZuHone10,ZuHone11a}. Synthetic microcalorimeter observations further show that the observed line-of-sight velocity gradient depends on viewing angle, with larger velocities expected when the sloshing plane is viewed closer to edge-on and smaller projected velocities for more face-on geometries \citep{ZuHone16}. The observed north-south velocity gradient in A2029 therefore suggests that the sloshing plane is not exactly in the plane of the sky, but has a finite inclination such that the northern side of the flow is moving partly toward the observer.

The observed velocity amplitude is modest and consistent with the sloshing interpretation. The largest line-of-sight contrast is $\Delta v_{\rm bulk} \sim 280~{\rm km\,s^{-1}}$, while the absolute bulk velocities are about $200~{\rm km\,s^{-1}}$ or less. Since this contrast is measured between opposite sides of the core, the characteristic one-sided projected velocity is of order $\sim100$--$150~{\rm km\,s^{-1}}$. For a true sloshing velocity of a few $100~{\rm km\,s^{-1}}$, such a projected amplitude can be produced if the sloshing plane is inclined by several tens of degrees from the plane of the sky. For example, $v_{\rm true}\sim300~{\rm km\,s^{-1}}$ gives $i\sim20$--$30^\circ$ for $v_{\rm los}=v_{\rm true}\sin i$. This estimate is illustrative, and tighter constraints would require tailored simulations. Thus, the observed velocity field can be understood as gentle, subsonic sloshing viewed at moderate inclination.

This interpretation is supported by the deep Chandra study of A2029 by \citet{Watson26}. They found that the large-scale spiral, southeastern splash feature, and possible northwestern shock are qualitatively reproduced by a $1{:}10$ mass-ratio off-axis merger with an impact parameter of 500 kpc, originally presented by \citet{ZuHone11a}. In this scenario, the observed structures correspond to an epoch about $0.2$~Gyr after the second core passage, or about $4.3$~Gyr after the start of the merger. The XRISM velocity gradient provides complementary dynamical evidence that the spiral structure traces sloshing-induced gas motions.

A similar picture has recently emerged from XRISM observations of the Centaurus cluster. In Centaurus, line-of-sight bulk velocities of $130$--$310~{\rm km\,s^{-1}}$ were detected within the central $\sim30$ kpc and interpreted as sloshing of the core gas, while the velocity dispersion is low, $\sigma_v\lesssim120~{\rm km\,s^{-1}}$, even near the central AGN \citep{XRISM25d}. A2029 shows a similar combination of ordered bulk motion and modest velocity dispersion, supporting the view that sloshing can produce measurable line-of-sight velocity gradients without generating large turbulence. The main difference is likely the viewing geometry and spatial scale: Centaurus appears to be viewed closer to edge-on with respect to the sloshing plane, whereas A2029 shows a large-scale spiral mostly in projection, with a finite but more modest line-of-sight component in the Resolve pointings.

\subsection{Non-thermal pressure support}
\label{sec:discussion_ntp}

The spatially resolved analysis allows us to test whether localized gas motions in the A2029 core introduce non-thermal pressure support missed by field-averaged measurements. Although the bulk velocity field shows a clear north-south gradient, the measured velocity dispersion is modest in all regions. As shown in Table~\ref{tab:spectral_results} and Figure~\ref{fig:par_map}, $\sigma_v$ is below about $150~{\rm km\,s^{-1}}$. For gas temperatures of $kT \sim 5$--$10~{\rm keV}$, this corresponds to a three-dimensional turbulent Mach number of only $\mathcal{M}_{\rm 3D} \sim 0.1$--$0.2$, giving a turbulent non-thermal pressure fraction, $\alpha_{\rm turb}$, of a few percent or less.

Including the ordered line-of-sight bulk velocity does not change this conclusion. When both $\sigma_v$ and the bulk velocity relative to the BCG are included in the effective kinetic-to-thermal pressure proxy, the resulting non-thermal pressure fraction $\alpha$ stays below about $2.5\%$ across the core. The largest value is found in the NW region, where both the velocity dispersion and the blueshifted bulk velocity contribute, while the center also shows a value of about $2\%$. Thus, even though the bulk velocities are detected at high significance, their contribution to the pressure budget is small.

This low kinetic pressure level can be compared with cosmological simulations. A forward-modeling comparison with the TNG-Cluster simulation suite \citep{Nelson24} by Lau et al. (submitted) shows that projection and azimuthal sampling can partly reduce the inferred turbulent pressure fraction, but that including ordered bulk motions still leaves A2029 below the simulated distribution. The discrepancy therefore appears to reflect the unusually low amplitudes of both turbulent and bulk motions in A2029. Thus, XRISM indicates that A2029 hosts sloshing-driven bulk flows while being dynamically quiescent in terms of turbulent pressure support.

This result provide a direct constraint on the turbulent contribution to hydrostatic mass bias in the central region of A2029. If non-thermal pressure support is ignored, the hydrostatic mass can be biased low by an amount comparable to the non-thermal pressure fraction, with additional dependence on its radial gradient. The present measurements show that, even after dividing the core into multiple regions and including ordered bulk motions, the kinetic pressure contribution is only a few percent, comparable to that inferred from the field-averaged XRISM measurements reported by \citet{XRISM25b}. Thus, the resolved velocity structure shows that the core is not kinematically featureless, but it does not imply a large departure from hydrostatic equilibrium. This conclusion is also consistent with the small hydrostatic-mass correction inferred by \citet{Bartalesi26}; although the XRISM data sets are not independent, the two analyses provide complementary views of the low level of kinetic pressure support in A2029.

The NE region requires some caution because its best-fit velocity dispersion is close to zero but the systematic uncertainty is not small (Appendix~\ref{app:systematics}). Since the line counts in this region are smaller than those in the bright center, the inferred intrinsic line width can be more sensitive to PSF spillover and statistical fluctuations. However, this does not affect the main conclusion that the data do not require a large turbulent pressure component. The derived non-thermal pressure fractions should therefore be regarded as estimates under the assumption that the measured velocity dispersion represents isotropic random motions. Within this framework, the current data do not indicate significant hydrostatic mass bias in the core of A2029.

\subsection{Implications for ICM heating and cooling}
\label{sec:discussion_heating}

The small non-thermal pressure fraction does not by itself imply that gas motions are irrelevant for the thermal balance of the cool core. Gas motions may dissipate kinetic energy and redistribute heat through mixing and sloshing-driven transport. We therefore estimate the turbulent heating rate and compare it with the radiative cooling rate in the core of A2029.

Following recent XRISM studies of the Perseus cluster \citep{XRISM26a}, we estimate the turbulent heating rate as
\begin{equation}
Q_{\rm heat} = C_0 \rho \frac{v^3}{l},
\label{eq:qheat}
\end{equation}
with $C_0=5$, using the measured line-of-sight velocity dispersion $\sigma_v$ as $v$ \citep{Zhuravleva14}. This assumes that the line broadening represents isotropic random motions, although unresolved bulk shear may also contribute. The coefficient $C_0$ already includes the three-dimensional isotropic correction. The radiative cooling rate is computed as
\begin{equation}
Q_{\rm cool} = n_{\rm e} n_{\rm i} \Lambda(T,Z),
\label{eq:qcool}
\end{equation}
using the X-COP electron density profile \citep{Eckert17,Ghirardini19}\footnote{\url{https://dominiqueeckert.wixsite.com/xcop/a2029}} and $\Lambda=2.5\times10^{-23}~{\rm erg~cm^3~s^{-1}}$, following \citet{XRISM26a}. For the length scale, we use the effective line-of-sight scale $l_{\rm eff}$ enclosing half of the emission measure \citep{Zhuravleva12,XRISM26a}. The density, density squared, and $l_{\rm eff}$ are averaged with projected emission-measure weights over 0.0--0.5 arcmin for the Center region, 0.5--1.5 arcmin for the NE, NW, SE, and SW regions, and 1.5--2.5 arcmin for the FNE region.  The $l_{\rm eff}$ values are $69.7_{-3.2}^{+3.3}~{\rm kpc}$ for the Center region, $109.0_{-0.8}^{+0.9}~\rm kpc$ for the NE, NW, SE, and SW regions, and $195.8\pm1.5~\rm kpc$ for the FNE region.

\begin{figure}
\begin{center}
\includegraphics[width=0.48\textwidth,clip,trim=0cm 0cm 0cm 1.0cm]{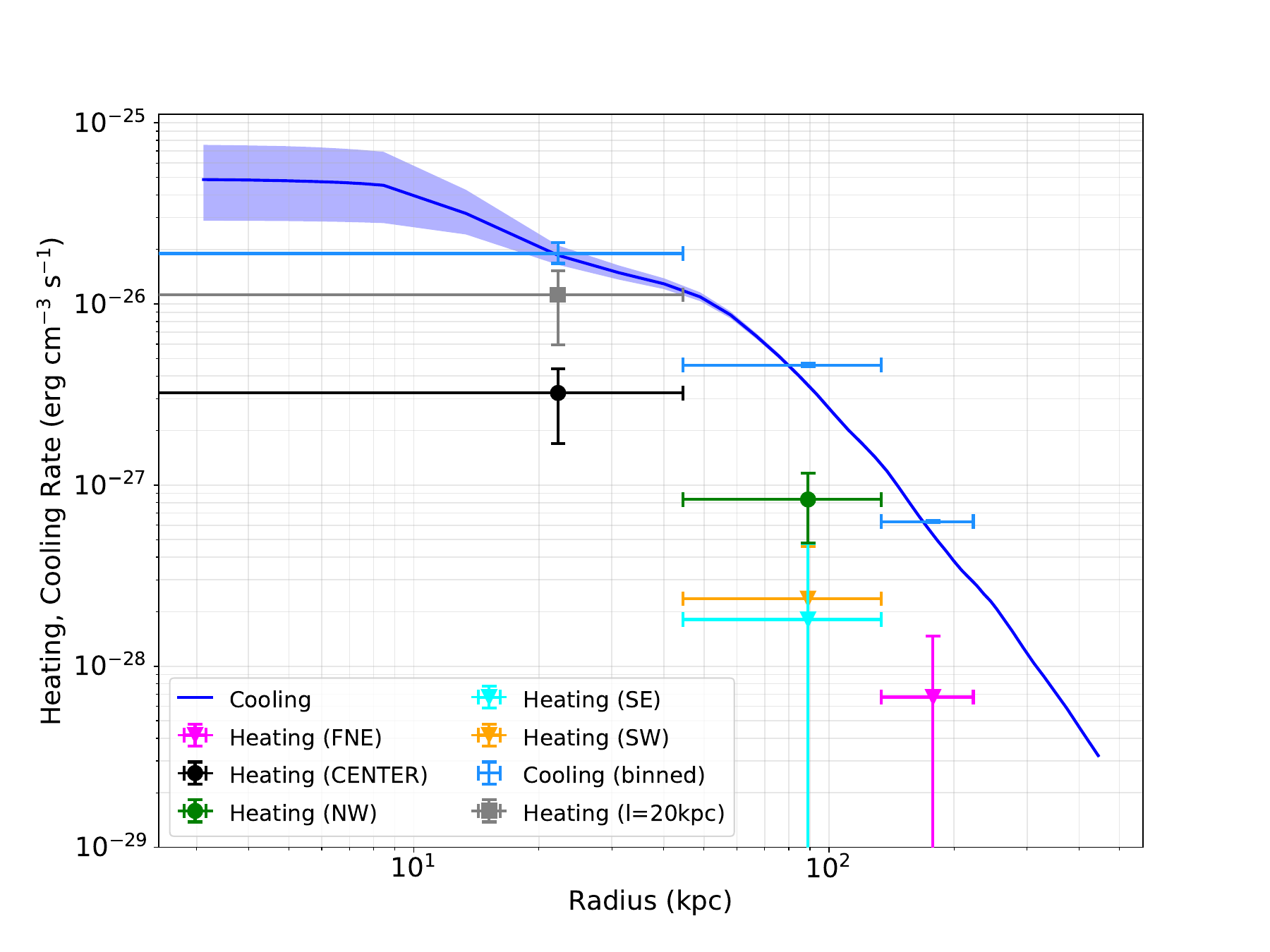}
\end{center}
\caption{Turbulent heating rate versus radiative cooling rate in the A2029 core (see Section~\ref{sec:discussion_heating}). The blue curve is the cooling rate from the X-COP density profile, with the shaded band showing the $1\sigma$ interval; sky-blue crosses show the same cooling rate binned into radial annuli. Symbols show $Q_{\rm heat}$ from the Resolve $\sigma_v$ values (Table~\ref{tab:spectral_results}): black and green circles for the Center and NW regions, and magenta, cyan, and orange downward triangles for FNE, SE, and SW, the latter close to upper limits. The NE point lies below the plotted range. Vertical error bars reflect the statistical uncertainties of $\sigma_v$.
The gray square point represents the heating rate estimated using injection length associated with AGN feedback, assuming radio/X-ray cavity size ($l=20\,{\rm kpc}$).
{Alt text: The figure compares turbulent heating rates with the radiative cooling rate as a function of radius. 
}}
\label{fig:heating_cooling}
\end{figure}

Figure~\ref{fig:heating_cooling} compares the turbulent heating rate with the radiative cooling rate. The cooling rate is computed from the X-COP density profile, while the heating rate uses the measured $\sigma_v$ values in Table~\ref{tab:spectral_results}. Because $Q_{\rm heat}\propto \sigma_v^3$, the statistical uncertainties are highly asymmetric where the intrinsic line width is weakly constrained. 

The estimated turbulent heating rate is lower than the radiative cooling rate in all regions when using $l_{\rm eff}$.
The largest ratios are found in the Center and NW regions, where $Q_{\rm heat}/Q_{\rm cool}$ is approximately 0.2. 
The ratio is approximately 0.1 in FNE, below 0.1 in SE and SW, and close to zero for the NE best-fit value because its velocity dispersion is consistent with zero. 
Using $l_{\rm eff}$ as the characteristic scale, turbulent dissipation alone appears insufficient to offset radiative cooling throughout the A2029 core. This conclusion is unchanged by excluding the Fe~\textsc{xxv} resonance line; the detailed test is described in Appendix~\ref{app:systematics}. In the central region, spatial scales associated with $\sigma_v$ could be lower than $l_{\rm eff}$ if motions are driven by the AGN feedback activity. Radio and X-ray observations of A2029 have revealed the presence of bubbles with the estimated sizes of $\sim 10-20$ kpc \citep{Timmerman2022}. Such scales bring heating rate closer to the cooling rate (gray point in Fig. \ref{fig:heating_cooling}). This cooling-heating trend is consistent with the results of gas density fluctuation analysis \citep{Zhuravleva18}.

This result differs from, but does not directory contradict,  the buoyancy-driven turbulent heating model presented in \citet{McNamara26}, primarily because the measured velocity amplitude here is slightly lower than than their adoped value taken from \citet{XRISM25a}, and the smaller characteristic scales predicted by their model.
Their calculation for A2029 uses the higher, field-averaged, single-temperature velocity dispersion of $165~{\rm km~s^{-1}}$ over the entire cooling volume and considers radio lobes rising at half the sound speed, where the sound speed is high because of the unusually high temperature of the A2029 atmosphere.
These assumptions lead to higher inferred turbulent heating rates than those obtained from our local velocity dispersions and $l_{\rm eff}$ values.
These optimistic assumptions used to test the model would imply injection scales of only a few kpc within 10 kpc and roughly 20 kpc at the 100 kpc cooling radius. These values lie far below $l_{\rm eff}$ and thus imply larger heating rates. Because the relevant driving scale is likely to vary spatially, with AGN bubbles affecting the central region and sloshing dominating much of the remaining cool core, a direct comparison with buoyancy-driven models requires a more detailed treatment than attempted here.
These and other issues are discussed in detail in \citet{McNamara26} and \citet{2025Rose}. 

The cooler gas components reported by \citet{Sarkar25}, including a $\sim2.85$ keV component in the core, may indicate that the thermal balance is not perfectly static, although this does not by itself imply an unimpeded classical cooling flow. The observed bulk-velocity pattern shows sloshing-driven gas motion, which can redistribute thermal energy through advection, mixing, and transport of higher-entropy gas from larger radii. Recent simulations by \citet{Fujita26} suggest that sloshing can either help suppress cooling or reduce the effectiveness of central heating, depending on how it redistributes dense core gas relative to the AGN heating region. 
In a related interpretation, \citet{Soker25} argued that the field-integrated XRISM velocity dispersion of A2029 is compatible with a mixing-heating scenario, in which turbulence promotes mixing of hot shocked jet material with the ambient ICM rather than directly supplying the full cooling losses. 
Our spatially resolved analysis separates sloshing-induced bulk motions from the modest unresolved velocity dispersion. Although these results do not directly test the mixing-heating scenario, gas motions may still contribute to thermal redistribution through mixing and sloshing-driven transport. The efficiency of these processes over the full cooling volume, however, remains uncertain.

\section{Summary}

We investigated the spatially resolved gas velocity structure in the central region of the relaxed cool-core cluster Abell 2029 using XRISM Resolve observations. By dividing the core into six regions (NE, NW, SE, SW, center, and FNE) and accounting for spatial-spectral mixing caused by the XRISM point-spread function, we measured the line-of-sight bulk velocities, velocity dispersions, and non-thermal pressure fractions of the ICM. Our main results are summarized as follows.

\begin{itemize}

\item We detect clear spatial variations in the line-of-sight bulk velocity across the cluster core. The northern regions are blueshifted relative to the brightest cluster galaxy (BCG), with the largest blueshifts of about $-190~{\rm km\,s^{-1}}$ in the NW and FNE regions, whereas the southern regions are close to zero or slightly redshifted, reaching about $+90~{\rm km\,s^{-1}}$ in the SE region. The largest north-south velocity contrast is approximately $280~{\rm km\,s^{-1}}$.

\item The observed velocity pattern is consistent with gas sloshing inferred from the spiral-like surface brightness and temperature structures seen in previous Chandra observations. The Resolve measurements provide spectroscopic evidence for ordered sloshing-induced gas motions in the core of A2029.

\item The velocity dispersion is highly subsonic, with measured values and upper limits below about $150~{\rm km\,s^{-1}}$. We find no clear evidence for strongly enhanced turbulence associated with the sloshing pattern, and this conclusion is not sensitive to the choice of data set or spectral modeling assumption.

\item The inferred non-thermal pressure fraction is below about $2.5\%$ in all regions, even when the measured bulk motions are included. Thus, departures from hydrostatic equilibrium due to gas motions are small in the observed core region.

\item Assuming that the measured velocity dispersion represents isotropic turbulence, the estimated average turbulent heating rate is below the radiative cooling rate. Turbulent dissipation alone is therefore insufficient to offset cooling, although gas motions may still contribute to thermal-energy redistribution through mixing and sloshing-driven transport. In the central region, the conclusion on the cooling-heating balance strongly depends on the size of the driving scale of turbulence.

\end{itemize}

These results show that the core of A2029 contains subsonic bulk flows naturally explained by sloshing, while the velocity dispersion and pressure support from gas motions are small. A2029 therefore provides an example of a relaxed cool-core cluster in which sloshing can be spectroscopically detected without substantially compromising the hydrostatic approximation.

\begin{ack}
We thank the XRISM project and the mission operations team for the operation of the satellite and their support of the observations. This work was supported in part by the Fund for the Promotion of Joint International Research, JSPS KAKENHI Grant Number 25K01026, 23H00121 (NO). NO acknowledges partial support by the Organization for the Promotion of Gender Equality at Nara Women's University. 
EDM acknowledges support from NASA grants 80NSSC25K7538 and 80NSSC24K0678.
SE acknowledges the financial contribution from {\it Theory Grant / Bando INAF per la Ricerca Fondamentale 2024} on ``Constraining the non-thermal pressure in galaxy clusters with high-resolution X-ray spectroscopy'' (1.05.24.05.10). BRM acknowledges the Canadian Space Agency for generous support.
\end{ack}

\appendix

\section{Systematic uncertainties in the gas velocity measurements}
\label{app:systematics}

We examined possible systematic effects on the spectral parameters, focusing on the gas velocity measurements. Since the line-of-sight velocity dispersion is only marginally constrained in some regions, we tested whether the results depend on the choice of data set, calibration uncertainty, treatment of spatial-spectral mixing caused by the broad point-spread function of XRISM, line-modeling effects such as resonant scattering, or thermal modeling. We also show the best-fit spectra for all regions to demonstrate the quality of the baseline fits.

\subsection*{Comparison among different data sets}

\begin{figure*}[tbh]
\begin{center}
\includegraphics[width=0.9\textwidth]{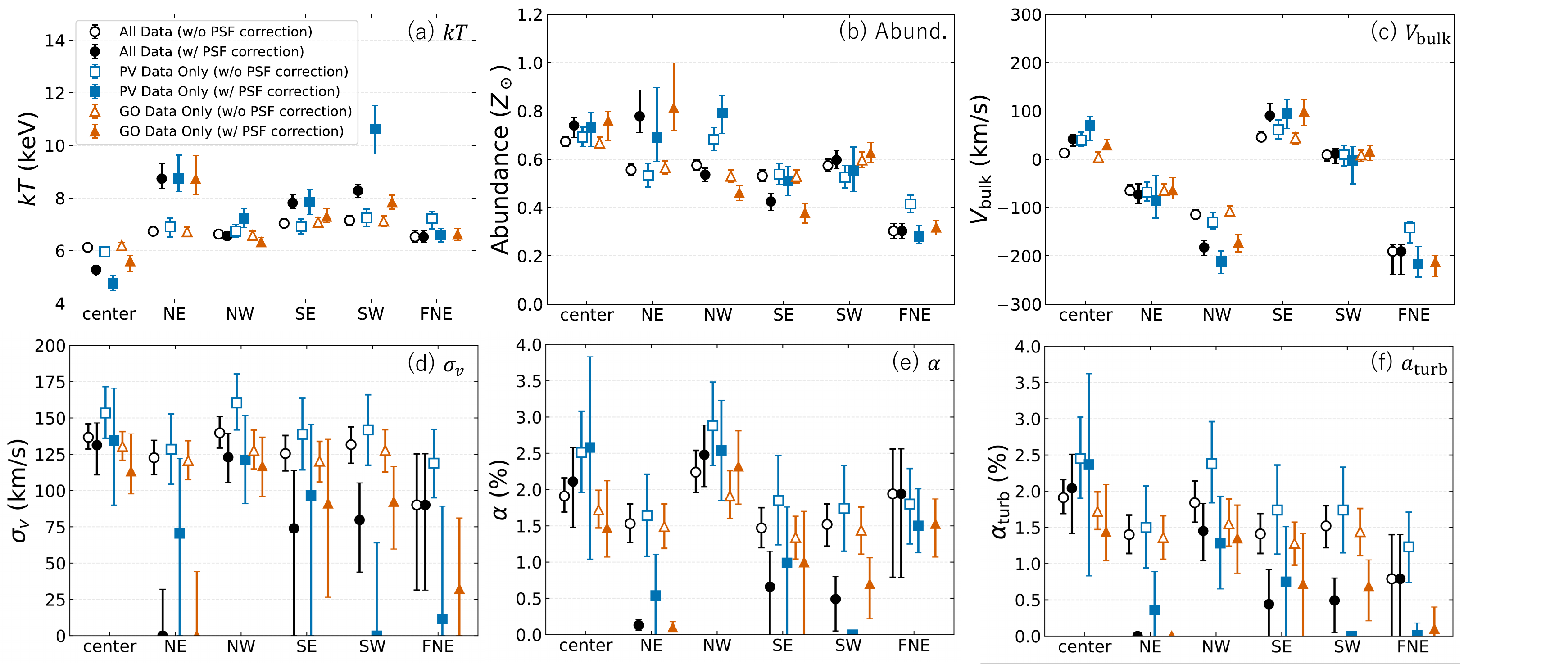}
\end{center}
\caption{Comparison of the best-fit spectral parameters obtained from the PV, GO, and combined PV+GO data sets. Panels show the temperature, metal abundance, bulk velocity relative to the BCG, velocity dispersion, and non-thermal pressure fraction. The results are generally consistent among the three data sets within the statistical uncertainties. 
{Alt text: Six panels compare spectral parameters from the PV, GO, and combined PV plus GO data sets for each region. The x-axis represents the source regions, and the y-axis gives the value of the corresponding parameter.
}}
\label{fig:app_dataset_comparison}
\end{figure*}

We compared the results obtained from the PV, GO, and combined PV+GO data sets. Figure~\ref{fig:app_dataset_comparison} shows that the temperature, metal abundance, bulk velocity, and velocity dispersion are generally consistent within the statistical uncertainties. In the PV-only analysis, the velocity dispersion in some regions tends to reach the lower boundary of the fit. When the GO data are included, the constraint no longer strongly favors the boundary solution, although it remains dominated by the upper limit. We therefore interpret the boundary solution in the PV-only fit as a consequence of limited photon statistics rather than evidence for an intrinsically vanishing line width. The combined PV+GO data provide the most stable constraints and are adopted as the baseline results.

\subsection*{Dependence on spectral modeling assumptions}

\begin{figure*}[tbh]
\begin{center}
\includegraphics[width=0.9\textwidth]{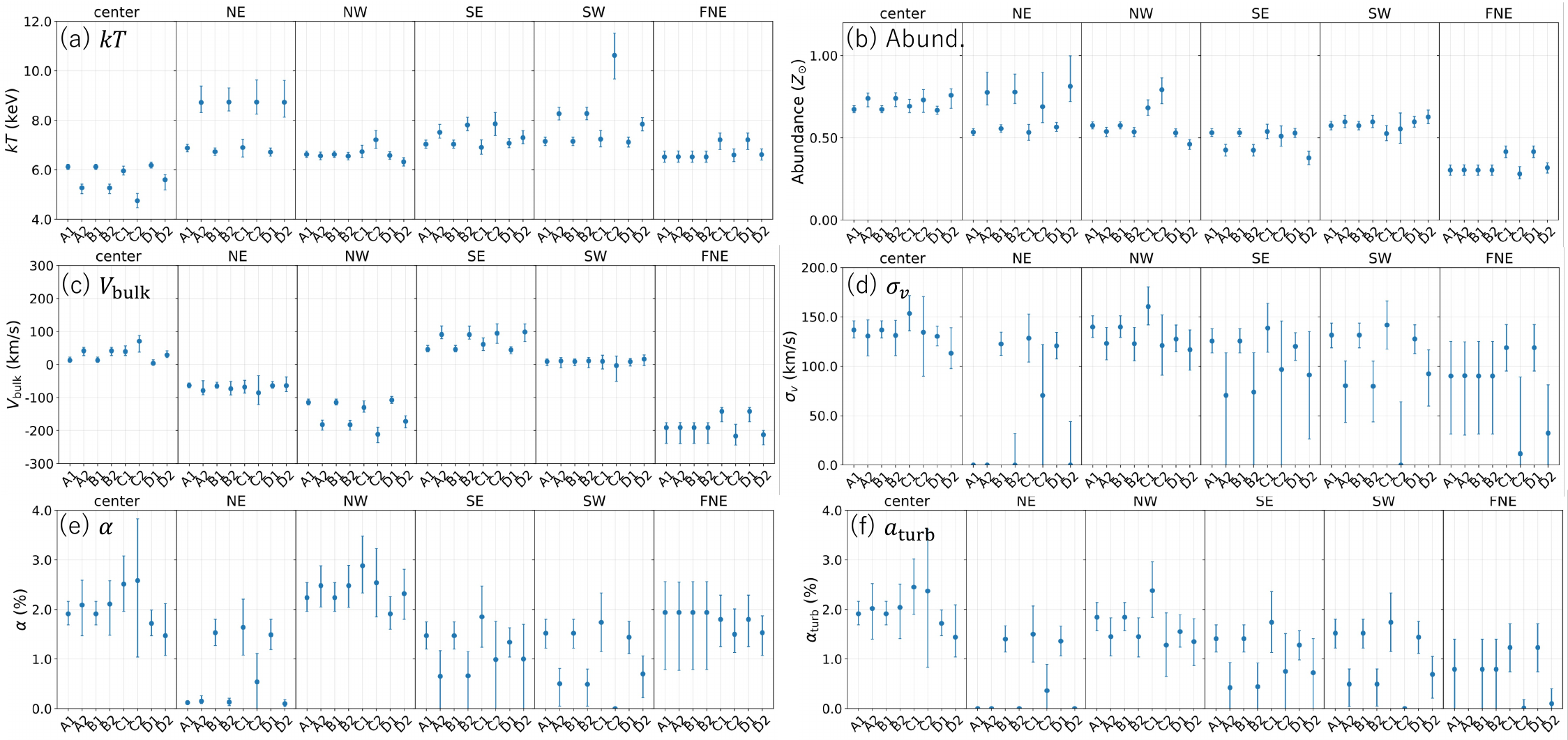}
\end{center}
\caption{Comparison of the best-fit spectral parameters for each region obtained under different spectral modeling assumptions. Panels show the systematic differences in the best-fit (a) temperature, (b) abundance, (c) bulk motion velocity, (d) velocity dispersion, (e) non-thermal pressure fraction, and (f) non-thermal fraction estimated from turbulent motions alone. The labels A, B, C, and D represent the analysis methods defined in the text: A fixes the NE velocity to zero, B allows all parameters except for $N_{\rm H}$ to vary, C uses only the PV data set, and D uses the GO data for the central region and the PV data for the northern region. The suffixes 1 and 2 denote analyses without and with SSM, respectively. 
{Alt text: Six panels compare the best-fit temperature, abundance, bulk velocity, velocity dispersion, total non-thermal pressure fraction, and turbulent non-thermal pressure fraction among the A1, A2, B1, B2, C1, C2, D1, and D2 analyses.
}}
\label{fig:app_model_comparison}
\end{figure*}

We also examined how the spectral parameters depend on the modeling assumptions. Figure~\ref{fig:app_model_comparison} compares several alternative fits, denoted as A1, A2, B1, B2, C1, C2, D1, and D2. In all these fits, the ICM emission was modeled with \texttt{TBabs*bapec}, the Galactic absorption column density was fixed at $N_{\rm H}=3.0\times10^{20}~{\rm cm^{-2}}$, the abundance table was set to \texttt{lpgs}, and the C-statistic was used. Case A fixes the velocity of the NE region to zero, motivated by the near-zero best-fit velocity in some fits. Case B allows all parameters except for $N_{\rm H}$ to vary. Case C uses only the PV data set. Case D uses the GO data for the central region and the PV data for the northern region. The suffix 1 denotes fits without SSM, in which each region is fitted independently, whereas suffix 2 denotes fits with SSM. Thus, A1 and A2 are fits with the NE velocity fixed to zero without and with SSM, respectively, B1 and B2 are fits with all parameters except for $N_{\rm H}$ free without and with SSM, respectively, C1 and C2 are the PV-only fits without and with SSM, respectively, and D1 and D2 are the GO-center plus PV-north fits without and with SSM, respectively.

Because the half-power diameter of the XRISM mirror is comparable to the size of the subregions analyzed in this work, photons from neighboring regions can contribute significantly to each extracted spectrum. We therefore adopt the SSM correction in the baseline analysis, in which each observed spectrum is represented as a linear combination of emission from all source regions weighted by the corresponding ancillary response files. As shown in Figure~\ref{fig:app_model_comparison}, the velocity parameters are generally consistent among the alternative models within the statistical uncertainties. Some thermodynamic parameters, especially the temperature and normalization, show larger model dependence, as expected from their sensitivity to photons scattered from neighboring regions and to the assumed thermal structure. These checks indicate that the SSM correction should be included, but it does not change the main conclusions on the gas velocity structure.

As an additional check of possible line-modeling effects, we repeated the spectral fits after excluding the Fe~\textsc{xxv} resonance line. The resulting redshifts were systematically shifted by about $\Delta z \sim -0.0001$, corresponding to a line-of-sight velocity shift of about $-30~{\rm km\,s^{-1}}$ at the redshift of A2029. The velocity dispersion decreased from $131$ to $83~{\rm km\,s^{-1}}$ in the Center region and from $123$ to $29~{\rm km\,s^{-1}}$ in the NW region, while it increased from $90$ to $128~{\rm km\,s^{-1}}$ in the FNE region. These changes indicate some dependence of the inferred $\sigma_v$ on the treatment of the strongest Fe line. Even for the increased FNE value, the turbulent heating rate remains below the radiative cooling rate. Thus, this diagnostic test does not change the qualitative north-south bulk-velocity pattern or the conclusion that turbulent heating alone is insufficient to offset cooling.

\subsection*{Comparison with Chandra temperatures and abundances}

As an additional check of the thermal modeling, we compared the XRISM temperatures and metal abundances with those obtained from Chandra spectra extracted from the same sky regions. The Chandra spectra were extracted from ObsID 4977 and analyzed with CIAO 4.15. Since the point-spread function of Chandra is much smaller than the region size, photon mixing between regions is negligible for this comparison. The spatial temperature trends are broadly consistent between XRISM and Chandra. The average XRISM-to-Chandra temperature ratio over the six regions is $0.94\pm0.14$, where the uncertainty represents the standard deviation among the regions. Although the Chandra temperatures are higher in four of the six regions, the offset is modest and does not indicate a problem in the XRISM spectral modeling. The metal abundances show a clearer offset, with an average XRISM-to-Chandra ratio of $0.82\pm0.20$, indicating that the XRISM abundances tend to be lower than the Chandra values. This trend is qualitatively consistent with recent International Astronomical Consortium for High Energy Calibration (IACHEC) studies of cluster cross-calibration\footnote{\url{https://iachec.org/wp-content/presentations/2026/IACHEC_Itsuki_Cluster_Cross-cal.pdf}}. We therefore do not regard the thermal-parameter differences as evidence for a problem in the velocity analysis. Fits with the temperature fixed to the Chandra values also give velocity parameters consistent with the baseline results within the statistical uncertainties.

\subsection*{Energy-scale and line-spread-function uncertainties}

We further considered calibration uncertainties in the energy scale and the line-spread function. The residual uncertainty in the energy scale around the Fe-K band corresponds to about 15 km s$^{-1}$ in bulk velocity, which is smaller than the statistical uncertainties of most regional velocity measurements and does not affect the detected north-south velocity pattern. The full width at half maximum (FWHM) estimated from the calibration data is typically 4.4--4.6 eV, with an uncertainty of about 0.15 eV. Its effect on the derived velocity dispersion is expected to be only a few km s$^{-1}$ and is therefore negligible for the present analysis. These calibration uncertainties are not included in the statistical error bars shown in the figures, but they do not alter the conclusions of this work.

\subsection*{Fit quality}

\begin{figure*}[tbh]
\begin{center}
\includegraphics[width=0.9\linewidth,angle=0,clip]{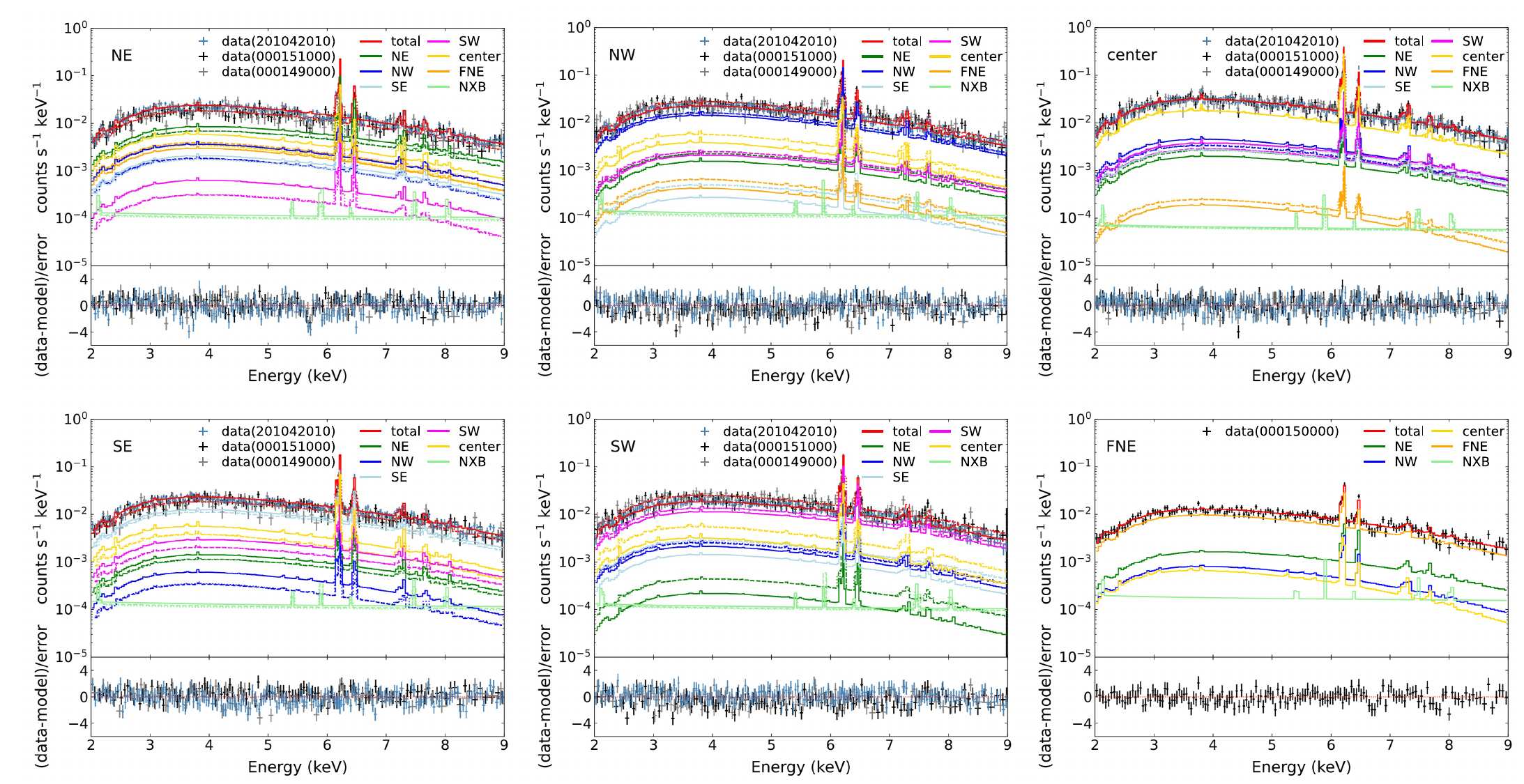}
\end{center}
\caption{Observed spectra and best-fit models for each region: (a) NE, (b) NW, (c) SE, (d) SW, (e) center, and (f) FNE. Crosses represent the observed spectra. The model includes emission components originating from all source regions and redistributed by the XRISM point-spread function, the non-X-ray background component, and their total. 
For NE, NW, center, SE and SW regions, the solid line represents the best-fit model for the GO observation, whereas the dashed and dotted lines represent the best-fit model for the two PV observations.
{Alt text: Six panels show the observed X-ray spectra and best-fit models for the NE, NW, SE, SW, center, and FNE regions. In each panel, the model consists of the total spectrum, point-spread-function redistributed emission from all regions, and the non-X-ray background component.}}
\label{fig:app_spectra_all}
\end{figure*}

\begin{figure*}[tbh]
\begin{center}
\includegraphics[width=0.9\linewidth,angle=0,clip]{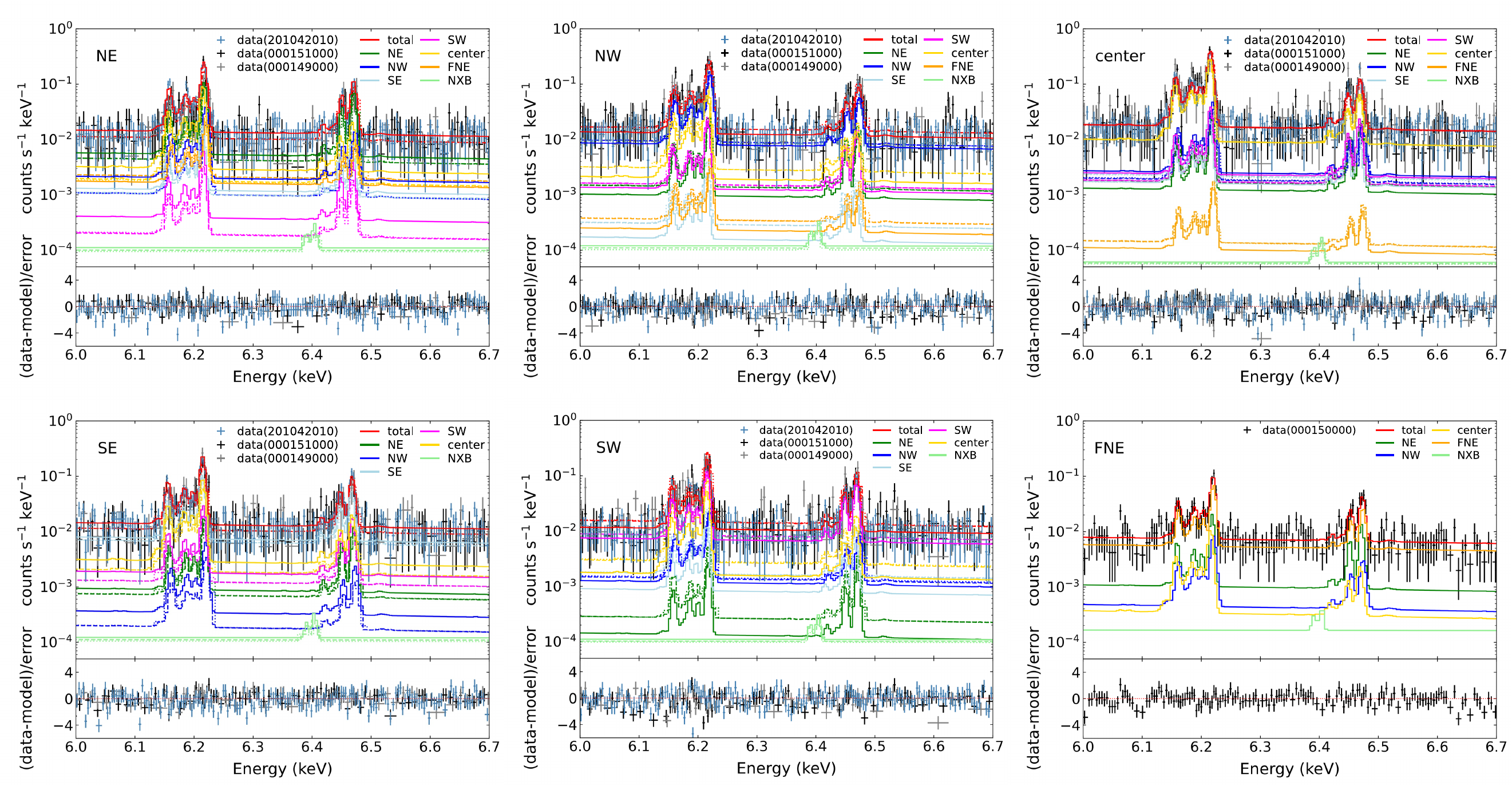}
\end{center}
\caption{Same as Figure \ref{fig:app_spectra_all}, but focusing on He-/H-like iron lines.
{Alt text: Six panels show the observed X-ray spectra and best-fit models for the NE, NW, SE, SW, center, and FNE regions focusing on He-/H-like iron lines.}}
\label{fig:app_spectra_H_He}
\end{figure*}

Figure~\ref{fig:app_spectra_all} shows the observed spectra and best-fit models for all regions. The spectra are well reproduced by the adopted model in the Fe-K band (Figure~\ref{fig:app_spectra_H_He}), where the gas redshift and velocity dispersion are mainly constrained. The model includes emission components originating from all source regions and redistributed by the XRISM point-spread function, as well as the non-X-ray background component. The good agreement supports the adequacy of the baseline spectral model for the velocity measurements.

\subsection*{Summary of the systematic checks}

The comparison among the PV, GO, and PV+GO data sets shows that the spectral parameters are mutually consistent and that the apparent zero-width solutions in some subsets are most likely caused by limited photon statistics. Alternative spectral models confirm that the SSM correction affects some thermodynamic parameters but does not change the main velocity results. The Fe~\textsc{xxv} resonance-line test suggests some sensitivity of $\sigma_v$ to line modeling, especially in the brighter regions, but does not change the qualitative bulk-velocity pattern. The Chandra comparison and calibration checks further support the robustness of the measured velocity structure. We therefore conclude that the low velocity dispersion and north-south bulk-velocity pattern are not driven by a particular data set, calibration uncertainty, or spectral modeling assumption.

\bibliographystyle{pasj}
\bibliography{refs}

\end{document}